\documentclass[11pt,a4paper]{article}
\usepackage[T1]{fontenc}
\usepackage{graphicx}
\usepackage{indentfirst}
\newcommand{\bi}[1]{\mbox{\boldmath$#1$}}
\newcommand{\balpha}{\mbox{\boldmath$\alpha$}}
\newcommand{\bomega}{\mbox{\boldmath$\omega$}}
\newcommand{\bmu}{\mbox{\boldmath$\mu$}}
\newcommand{\bsigma}{\mbox{\boldmath$\sigma$}}

\begin{document}
\begin{center} 
\huge{\bf Classical elementary particles, spin, \\{\it zitterbewegung} and all that}
\end{center}
\vspace{0.5cm}

\begin{center}
\large{\bf Mart\'{\i}n Rivas}

\vspace{0.5cm}

\large{{Dpto. de F\'{\i}sica Te\'orica, Universidad del Pa\'{\i}s Vasco-Euskal Herriko Unibertsitatea,\\ 
Apdo.~644, 48080 Bilbao, Spain}\\{e-mail: wtpripem@lg.ehu.es}}
\end{center}

\vspace{0.5cm}

\noindent{\bf Abstract}
After a revision of the main features of the structure of the Dirac electron
a plausible definition of elementary particle is stated. It is shown that this definition
leads in the classical case to a picture which produces a very clear 
correspondence between the classical
and quantum mechanical features of the electron. It is analyzed how the classical spin
structure and {\it zitterbewegung} are related to the classical variables that define
the kinematical state of the particle.

\vspace{2cm}

\section{Introduction}
\label{sec:intro}

The very concept of elementary particle seems to be not sufficiently 
well stated in undergraduate physics books. In most cases they are mentioned
as the bricks with which matter can be built. But, if they are so simple
objects it is important to establish the theoretical difference between them
and other small objects which are not considered as elementary.
One of the goals of this work is to introduce
some simple ideas which shed light about a theoretical 
concept of elementary particle and how
these ideas lead to their classical and quantum mechanical description.

We know that matter is formed by smaller and smaller 
units which can be divided again and again into smaller 
pieces. This process can be continued until we hypothetically
reach what the Greeks called an atom, the indivisible unit of matter. 
Today we call this last
division of matter an elementary particle. 
Elementary particles should be thus the building blocks with which
all material systems can be formed.

Leptons and quarks are the basic elementary particles in the standard model picture.
They all have, as intrinsic properties, mass, spin 1/2, electric charge 
and several other quantum numbers with some exotic names as baryon and lepton number, 
isotopic spin, strangeness, beauty and others.
Together the intermediate spin 1 bosons, which carry the interactions between them,
we have a collection of, charged and uncharged, spinning objects. It seems that there 
are no spinless elementary particles in nature, 
so that the spin description is crucial for the understanding 
of the structure of matter.
Whether or not these todays considered elementary particles can be subsequently
divided is still an open question. But if these final indivisible objects exist,
it is possible to state, at least from the theoretical point of view, 
what are the necessary conditions for a mechanical system to be considered
as elementary, to be considered as the ending process of division of matter.

To fix ideas we review in the next section, the description, from the classical 
and quantum mechanical point of view, of the structure of the electron.

We state in section \ref{concept} the concept of elementary particle as a mechanical
system without excited states. The description of the state of an elementary particle is reduced
to the analysis of the updated consecutive inertial observers which describe the particle in the same state.
The remaining sections are devoted to the analysis of the 
kind of Lagrangian systems the above definition allows to deal with, in particular 
with the relativistic and non-relativistic point particle 
and more important, of the spinning systems.
We end this work with the classical Lagrangian description of the photon and a final section
devoted to general conclusions and some difficulties of the formalism.

\section{What is an electron?}

The usual attempt to describe classically the spinning 
electron has been to endow the Newtonian
point particle with some spherical shape of small radius, 
some plausible charge distribution on its surface or in its whole volume,
and a final rotation as a rigid body to give account of the spin
angular momentum. 
Nevertheless this rigid body description has led to serious difficulties.
One needs to justify how the charge is held together and therefore
the additional description of the glue forces is necessary. With a finite radius
and a uniform mass distribution the velocity of the outer part of the sphere
should exceed the velocity of light to obtain the measured values of the spin.

The failure of this kind of classical models 
led Barut to quote about the structure of the spinning electron: 
{\it `If a spinning particle is not quite a point particle, nor
a solid three dimensional top, what can it be?
What is the structure which can appear under probing with
electromagnetic fields as a point charge, yet as far as spin and
wave properties are concerned exhibits a size of the order of
the Compton wavelength?'} \cite{Barut}.

Nevertheless, for the description of the electron what we 
have is the quantum mechanical Dirac's analysis.
Dirac equation has as many negative energy solutions as positive energy ones,
and one of the success of this formalism is the interpretation of the 
negative energy solutions as representing positive energy solutions
of a particle of the same intrinsic attributes mass and spin 1/2 but opposite
electric charge. Every charged spin 1/2 particle has an antiparticle.

If point ${\bi r}$ is the position vector on which Dirac's spinor 
$\psi(t,{\bi r})$ is defined, then, when analyzing the velocity of point 
${\bi r}$, Dirac finds that: \cite{Dirac}
\begin{enumerate}
\item{The velocity ${\bi v}=i/\hbar[H,{\bi r}]=c\balpha$, is expressed in terms of 
$\balpha$ matrices, which have eigenvalues $\pm1$, 
and the velocity of light $c$, and writes:\\ {\it 
`$\ldots$ a measurement of a component of the velocity of a free 
electron is certain to lead to the result $\pm c$. 
This conclusion is easily seen to hold also when there is a field present.'}}

\item{One of the puzzling features of this description is that
the linear momentum ${\bi p}$ does not have the direction of the velocity 
${\bi v}$ of point ${\bi r}$, but it is related to some average value of it: ${\ldots}$ 
{\it `the $x_1$ component of the velocity, $c\alpha_1$, consists of 
two parts, a constant part $c^2p_1H^{-1}$, connected with the momentum 
by the classical relativistic formula, and an oscillatory part, whose 
frequency is at least $2mc^2/h$, ${\ldots}$'}. \\
It seems that point ${\bi r}$ does not represent the center of mass position
of the electron.}

\item{The position ${\bi r}$ oscillates very fast in a small region of order of
Compton's wavelength:\\ {\it `The oscillatory part of $x_1$ is 
small, ${\ldots}$ , which is of order of magnitude $\hbar/mc$, 
${\ldots}$'}.\\
This oscillatory motion of point ${\bi r}$ was defined by Schr\"odinger
as the {\it zitterbewegung} or quivering or trembling motion of the electron.}

\item{In his Nobel dissertation lecture he insists on this peculiar feature of the 
quivering motion at the speed of light
of the electron: \cite{DiracNobel}\\ `{\it The new variables $\balpha$, 
which we have to introduce to get a relativistic wave
equation linear in $H$, give rise to the spin of the electron. From the general
principles of quantum mechanics one can easily deduce that these variables $\balpha$
give the electron a spin angular momentum of half a quantum and a magnetic
moment of one Bohr magneton in the reverse direction to the angular
momentum. These results are in agreement with experiment. They were, in
fact, first obtained from the experimental evidence provided by spectroscopy
and afterwards confirmed by the theory.'}\\
`{\it The variables $\balpha$ also give rise to some rather unexpected phenomena concerning
the motion of the electron. These have been fully worked out by
Schr\"odinger. It is found that an electron which seems to us to be moving
slowly, must actually have a very high frequency oscillatory motion of small
amplitude superposed on the regular motion which appears to us. As a result
of this oscillatory motion, the velocity of the electron at any time equals the
velocity of light. This is a prediction which cannot be directly verified by
experiment, since the frequency of the oscillatory motion is so high and its
amplitude is so small. But one must believe in this consequence of the theory,
since other consequences of the theory which are inseparably bound up with
this one, such as the law of scattering of light by an electron, are confirmed
by experiment.'}\\
The electron has thus a {\it regular motion}, which can be 
easily interpreted as the motion
of its center of mass, and the oscillatory motion of point ${\bi r}$ at the velocity of light
around the center of mass. The absolute value of the velocity $c$ is not modified by the presence of an
external field.}

\item{The total angular momentum of a free electron ${\bi J}={\bi r}\times{\bi P}+{\bi S}={\bi L}+{\bi S}$, 
is a constant of the motion, but the orbital part ${\bi L}$ and the spin ${\bi S}$ are not separately
constants of the motion.
The spin ${\bi S}$ of a free electron satisfies the dynamical equation
\[
\frac{d{\bi S}}{dt}={\bi p}\times c{\balpha}={\bi p}\times{\bi v},
\]
so that the spin operator ${\bi S}$ is only a constant of the motion for the center of mass observer for which
${\bi p}=0$.
}

\item{In his original 1928 paper \cite{Diracp}, when analyzing the 
interaction of the electron with an external electromagnetic field, 
after developing the interaction Hamiltonian to first order in the external fields, 
he obtains two new interaction terms: 
 \begin{equation}
{e\hbar\over 2m}{\bf\Sigma}\cdot{\bi B}+{ie\hbar\over 2mc}\balpha\cdot{\bi E}=-\bmu\cdot{\bi B}-{\bi d}\cdot{\bi E},
 \label{eq:D8}
 \end{equation}
where Dirac's electron spin operator is written as \[
{\bi S}=\frac{\hbar}{2}{\bf\Sigma}=\frac{\hbar}{2}\pmatrix{\bsigma&0\cr 0&\bsigma\cr},
\] 
in terms of $\sigma$-Pauli matrices and where ${\bi E}$ 
and ${\bi B}$ are the external electric and magnetic fields, 
respectively. He says: {\it `the electron will therefore behave as 
though it has a magnetic moment $-(e\hbar/2m)\,{\bf\Sigma}$ and an 
electric moment $-(ie\hbar/2mc)\,\balpha$. The magnetic moment
is just that assumed in the spinning electron model'}. `{\it The 
electric moment, being a pure imaginary, we should not expect to 
appear in the model. It is doubtful whether the electric moment has any physical meaning.'} 
}

\end{enumerate}

The electron, in addition to the electric charge behaves as though 
it has some electric and magnetic dipole moments. The magnetic dipole term is the right
one to give account of the Zeeman effect in atoms. But, what about the electric dipole? 
In the last Dirac sentence it is difficult to understand why Dirac, who did not reject the
negative energy solutions, and therefore its consideration as the antiparticle states,
and insisted that the motion of point ${\bi r}$ is at the speed of light as 
an {\it `inseparably bound up'}
consequence, disliked the existence of this electric dipole
which was obtained from his formalism on an equal footing 
and at the same time as the magnetic dipole term. In his book and in the
Noble dissertation he never mentioned again this electric dipole property.
But, what happens if the point ${\bi r}$ represents the position of the center of charge of the electron.
By the previous analysis it seems to be a different point than the center of mass.
Then this separation implies that for the center of mass observer there is a nonvanishing
electric dipole moment. It is oscillating very fast, its average value is probably 
zero, but it is a physical property that has to be taken into account. 

In fact, in quantum electrodynamics, the complete Dirac Hamiltonian
contains both terms, perhaps in an involved way because the above expression (\ref{eq:D8})
is a first order expansion in the external fields considered as classical commuting fields. 
It might happen that this electric dipole does not represent the existence of a particular 
positive and negative charge distribution for the electron, but rather a separation
between the center of mass and center of charge. This interpretation will be obtained later
when analyzing the classical spinning models.

\section{What is an elementary particle?}
\label{concept}

The concept of elementary particle rests on the idea that this ultimate indivisible object
is a so simple mechanical system that it has no excited states. 
In a broad sense it is always 
in a single state which looks different for the different inertial observers, 
but the difference is only a matter of change of point of view and 
not a change of its intrinsic properties. It is only a kinematical change.

In the quantum case, once a single inertial observer $O$ describes the state of the electron, 
the collection of the remaining states described by all other inertial observers
completely reproduce the Hilbert space of all states. These states 
can be obtained from the previous one by the corresponding transformation
of the wave function to the new frames. 
There is no other possible state of the electron which cannot be described by any one 
of the above states
or any linear combination of them. 
The Hilbert space is the kinematical state space of an elementary particle. 
This set of states contains only kinematical modifications of any one
of them.

This is Wigner's quantum definition of an elementary particle as a system whose
Hilbert space of pure states is an invariant space under the group of space-time
transformations. 
In group theoretical language, an elementary particle is thus a quantum system
whose Hilbert space of states carries an irreducible representation of the kinematical
group of space-time transformations. \cite{Wigner}

It is clear that Wigner's irreducibility condition is a necessary condition for a
quantum system to be considered as elementary. But, what is the classical
equivalent requirement?

Let us assume that a particular inertial observer is describing the
classical state of a particle by giving the values 
of the basic and essential variables which characterize its state.
Now the particle, under some external influence, changes some of its
variables and therefore it goes into a different state. 
Then, by assumption, if the particle is elementary
it is always possible to select another inertial observer
who describes the particle in the same state and with the same values of its
variables as the previous observer. This is the idea about what an elementary
particle should be from a theoretical viewpoint. 
Every change in the state of an elementary particle
can always be compensated by choosing a new inertial reference frame.

This idea imposes very stringent conditions on what are the classical variables
we need to describe the states of an elementary particle.
For instance, if the particle has some directional property like spin and the spin orientation changes,
the new inertial observer should accommodate its reference frame to describe the spin
components as before and therefore its frame looks rotated with respect to the previous one.
By following the change of orientation of the updated inertial frames, 
we are describing the change of orientation
of the spin of the particle, and thus the spin dynamics.

At this stage we do not know yet 
what are the classical variables to describe an elementary particle. But instead of describing
the particle let us try to describe the collection of consecutive updated inertial observers
who instantaneously describe the elementary particle by the same state. By describing every one
of these particular inertial observers we are describing the instantaneous 
kinematical state of the elementary particle.
By describing the instantaneous change of these inertial observers we are describing the dynamical 
behavior of the particle.

Let us see first how we use to describe every reference inertial frame with respect to each other.
This is very deeply related to a restricted relativity principle.

\section{The Restricted Relativity Principle}

A very fundamental principle is a Restricted (sometimes called Special) relativity principle. 
By restricted it is meant that the fundamental laws of physics have to be invariant for a restricted
set of equivalent observers, called inertial observers. 
It is also called restricted to distinguish
from a General relativity principle in which all observers, accelerated or not, 
describe nature by the same form invariant laws.

The relationship between the different inertial observers is related to their relative measurement
of space-time events, {\it i.e.}, to how they relate the time and position coordinates
of the same and every space-time event. This relationship is defined by a 
space-time transformation group which is called a kinematical group.
Different kinematical groups containing as possible transformations space and time translations, 
rotations and boosts, 
are analyzed by Bacry and Levy-Leblond in \cite{Bacry}. 
Among them we find the Galilei and Poincar\'e group
and also the de Sitter groups. 
But more general groups, like the conformal group, could be good candidates as kinematical groups.
If we restrict ourselves to the Galilei and Poincar\'e group we shall obtain only as intrinsic attributes
of matter the value of the mass and spin, which are the two group invariant properties. Other
quantum numbers have no so clear classical interpretation and are related to the so called `internal'
symmetry groups.

Let us consider first the Galilei group. 
The relationship between two arbitrary inertial observers is:
\[
t'=t+b,\quad {\bi r}'=R(\balpha){\bi r}+{\bi v}t+{\bi d}.
\]
Observer $O$ measures the coordinates $t$ and ${\bi r}$ of a particular space-time event 
while observer $O'$ obtains the coordinates $t'$ and ${\bi r}'$ for the same event. Then the remaining
variables of the above formulae $b$, ${\bi d}$, ${\bi v}$ and $R(\balpha)$, 
which are constant once the two inertial observers are fixed,
define the relative relationship between them. Any relative measurement 
between them of any other physical property of a system will depend on these constants.
Let us see their meaning. 

Suppose observer $O$ measures the emission of a burst of light from the origin
of its Cartesian frame ${\bi r}=0$ when its clock shows $t=0$. This event takes place
for $O'$ at time $t'=b$ and at the point ${\bi r}'={\bi d}$. If at instant $dt$ observer 
$O$ measures another light burst form its origin, it takes the values $t'+dt'=dt+b$
and ${\bi r}'+d{\bi r}'={\bi v}dt+{\bi d}$, {\it i.e.}, after a time $dt'=dt$ and displaced 
$d{\bi r}'={\bi v}dt$ with respect to the previous event. 
Therefore the origin of $O$ is moving with a velocity $d{\bi r}'/dt'={\bi v}$
as measured by $O'$.
Finally $R(\balpha)$ represents how the Cartesian unit vectors of $O$ are represented in $O'$ frame
if they were at relative rest.

Then $O'$ describes observer $O$ by giving at its time $b$ the position of the origin of $O$
frame at ${\bi d}$, moving with velocity ${\bi v}$ and with three unit vectors linked to it
and oriented according to $R(\balpha)$. We say that $O'$ specifies any other inertial observer
by some fixed values of the time, the position of a point, velocity of that point and orientation around that
point. That's all.

\section{Classical Elementary Particles}

Let us think now that we consider ourselves identified with the observer $O'$ 
and we are measuring the evolution of an 
elementary particle in terms of some arbitrary evolution parameter $\tau$, which can be 
our own time or any other monotonic function of it. Now let us focus our attention to some
particular inertial observer. 
It can be, for instance, an inertial observer very close to the particle or an inertial observer
located in some specific part of the system. 
Then, at instant $\tau$,
we fix this particular inertial observer by giving the 
ten variables $(t(\tau),{\bi r}(\tau),{\bi v}(\tau),\balpha(\tau))$, 
with the same meaning respectively as before, although we have partially changed the notation.
The time translation parameter $b$ has been replaced by the time observable $t(\tau)$ and the position
of the origin ${\bi d}$ by ${\bi r}(\tau)$. 
By $\balpha(\tau)$
we mean the three parameters which can be used to describe the orientation of any frame. 
They can be Euler's angles or any other parametrization of the rotation group. In any case we give
the values of a time $t$, the position of a point ${\bi r}$, the velocity
of this point ${\bi v}$ and the orientation around this point ${\balpha}$ or 
the matrix of the three unit vectors $R(\balpha)$.

Now at instant $\tau+d\tau$ 
the particle has changed, so that 
the new selected inertial observer who sees the particle in the same state as the previous 
inertial observer, is described by the new variables $(t(\tau+d\tau),{\bi r}(\tau+d\tau),
{\bi v}(\tau+d\tau),\balpha(\tau+d\tau))$.
The group structure of the Galilei group implies 
that the relationship between these two inertial observers is given by an infinitesimal element
of the Galilei group $\delta g$ of parameters $(dt,d{\bi r},d{\bi v},d\balpha)$, so that the change
of the action of the system between instants $\tau$ and $\tau+d\tau$ is given to first order
in terms of these variations by
\[
Ld\tau=Tdt+{\bi R}\cdot d{\bi r}+{\bi V}\cdot d{\bi v}+{\bi A}\cdot d\balpha=
(T\dot{t}+{\bi R}\cdot \dot{\bi r}+{\bi V}\cdot\dot{\bi v}+{\bi A}\cdot\dot{\balpha})d\tau,
\]
where the overdot means derivation with respect to the evolution parameter $\tau$. We do not know
yet the functions $T$, ${\bi R}$, ${\bi V}$ and ${\bi A}$, but what is clear is that the 
most general Lagrangian which describes this system must be a function of the kinematical variables 
$(t,{\bi r},{\bi v},\balpha)$ and their first order $\tau$ derivatives.

The evolution of the system in the interval $[\tau_1,\tau_2]$ is described, once the Lagrangian
is known, by giving these kinematical variables $(t,{\bi r},{\bi v},\balpha)\equiv x(\tau_1)$ at instant
$\tau_1$ and the values of the same variables at the final instant $\tau_2$, $x(\tau_2)$.
The variational principle implies that the path followed by the 
system between these fixed end points $x(\tau_1)$ and $x(\tau_2)$, 
is the one which produces a minimum
for the action of the system. The kinematical or evolution space
of this Lagrangian system is just the manifold spanned by these initial (or final) points.
But, because any infinitesimal change is produced by an infinitesimal element of the Galilei
group $\delta g$, 
the composition of all these consecutive infinitesimal group elements will produce 
in general a finite group element $g$ such that $x(\tau_2)=gx(\tau_1)$. The manifold
spanned by these variables has the property that given any two points there exists
at least a group element $g$ that relates them. In group theory it is said that 
the group acts transitively on this manifold or that the manifold is a homogeneous space of the group.

We conclude that the kinematical space of a classical elementary particle must 
necessarily be a {\it homogeneous space} of the kinematical group. 
This is the classical requirement equivalent to the irreducibility
condition of the quantum case to define a classical elementary particle.

Although we started this analysis with the Galilei group we see that the parametrization of
the Poincar\'e group leads to the same kind of ten variables $(t,{\bi r},{\bi v},\balpha)$,
and with the same geometrical meaning as before, to characterize the relative description
of every inertial observer. This implies that, even in the relativistic case, the most general
Lagrangian of a classical elementary particle can be written as before as
 \begin{equation}
L=T\dot{t}+{\bi R}\cdot \dot{\bi r}+{\bi V}\cdot\dot{\bi v}+{\bi A}\cdot\dot{\balpha}.
 \label{eq:gen}
 \end{equation}

As a matter of fact we see that the Newtonian point particle is a classical elementary particle,
according to this definition.
The kinematical space is spanned by $t$ and ${\bi r}$, 
which is a homogeneous space of the Galilei group. 
Because the free motion is at a constant speed
and the particle has no orientation properties, 
it is not necessary to modify neither the velocity ${\bi v}$
nor the orientation $\balpha$ of the successive inertial observers, so that the variation of the action
is written as
\[
Ld\tau=(T\dot{t}+{\bi R}\cdot\dot{\bi r})d\tau,
\]
and the most general free Lagrangian for a point particle is a function of $t$, ${\bi r}$ and their 
first order $\tau-$derivatives. Variables $t$ and ${\bi r}$ are interpreted as the time and 
position observable of the particle, respectively, and in a time evolution description the Lagrangian 
will be also a function of the velocity $d{\bi r}/dt$. The above expansion suggests that
the Lagrangian can be written as
 \begin{equation}
L=T\dot{t}+{\bi R}\cdot\dot{\bi r}.
 \label{eq:point}
 \end{equation}
If the Lagrangian of any Lagrangian system is written in terms of the kinematical variables, {\it i.e.,}
in terms of the end point variables $x_i$ of the variational formalism, 
instead of the independent degrees of freedom,
then the Lagrangian is a homogeneous function of first degree of the $\tau-$derivatives 
of the kinematical variables $\dot{x}_i$. Therefore, Euler's theorem on homogenous functions
of first degree in $\dot{x}_i$ imply
\[
L(x,\dot{x})=\frac{\partial L}{\partial\dot{x}_i}\,\dot{x}_i,
\]
with the usual addition convention on repeated indexes.
This homogeneity property can be seen for instance in \cite{Rivasl} in which a detailed
analysis of this property, even for generalized Lagrangians depending on higher order derivatives, 
is worked out. 

Then the above expansion (\ref{eq:point}) 
is just this homogeneity condition and thus
\[
T=\frac{\partial L}{\partial\dot{t}},\quad {\bi R}=\frac{\partial L}{\partial\dot{\bi r}}.
\]
In the $\tau-$evolution description we see in the non-relativistic case
\[
L dt=L\dot{t}\,d\tau=\frac{m}{2}\left(\frac{d{\bi r}}{dt}\right)^2\dot{t}\,d\tau=\frac{m}{2}\frac{{\dot{\bi r}}^2}{\dot{t}} d\tau\equiv \widehat{L}(x,\dot{x})d\tau
\]
{\it i.e.,} $\widehat{L}(x,\dot{x})=m{\dot{\bi r}}^2/2\dot{t}$, a homogeneous function of first degree in terms of $\dot{t}$
and $\dot{\bi r}$. 

Similarly, in the relativistic case, the space-time manifold is also a homogeneous space 
of the Poincar\'e group. We also get for the point particle Lagrangian a homogenous function
of first degree in terms of the derivatives of the kinematical variables
\[
\widehat{L}(x,\dot{x})=-mc\sqrt{c^2{\dot{t}}^2-{\dot{\bi r}}^2},
\]
since it is the square root of a second order polynomial in terms of the $\dot{x}$.

When compared the point particle Lagrangian (\ref{eq:point}) 
with the most general Lagrangian (\ref{eq:gen}), we see that new possible terms depending on $\dot{\bi v}$
and $\dot{\balpha}$ should appear. It is from the dependence on these new variables 
that the spin structure of the particle and the {\it zitterbewegung} of point ${\bi r}$ will emerge.

\section{Spinning Elementary Particles}

We propose the name of kinematical theory of elementary particles for this formalism because
the kinematical space of the system is basically related, as a homogeneous space, 
to the kinematical group of space-time transformations. Not only the kinematical
group defines the space-time symmetries of the system, but it also supplies 
the classical variables which characterize the description of a classical elementary particle.
It is the kinematical group 
of space-time transformations which defines the structure of the Lagrangian of an elementary particle.
We remit the reader to the published research works on this subject for the different technical points
and limit ourselves to outline the main features of the formalism \cite{Rivasp}. 

One technical thing is that the invariance of the dynamical equations under the kinematical group
of transformations does not mean that the Lagrangian has to be invariant. The invariance or 
not of the Lagrangian is related to the group structure \cite{LLCMP}. In the non-relativistic case
the Lagrangian transforms under the Galilei group up to a total $\tau-$derivative, 
while it is invariant under the Poincar\'e
group, provided the kinematical variables that define the mechanical system span 
a homogeneous space of the group. If these kinematical variables do not span a homogeneous space
the Lagrangians are no longer invariant, as it happens for compound systems.

The most general particle has thus ten kinematical variables $x\equiv(t,{\bi r},{\bi v},\balpha)$, 
which correspond to a mechanical system of six degrees of freedom. 
Three represent the position of a point ${\bi r}$
and the other three $\balpha$, its orientation in space like a rigid body. 
But, since the Lagrangian also depends on $\dot{x}$ it will depend on 
the derivative of ${\bi v}$, and thus the second derivative
of ${\bi r}$ and on the first derivative of $\balpha$.
Since the most general Lagrangian for an elementary particle depends on the
acceleration of point ${\bi r}$, the dynamical equations of this point will be of fourth order.
Analysis and description of the relativistic and non-relativistic fourth order differential 
equations to describe the spinning electron
are obtained in \cite{Rivaselec}. 

Instead of the dependence of the Lagrangian on the derivative of the orientation $\dot{\balpha}$
is better to use the angular velocity $\bomega$, which is a linear function of $\dot{\balpha}$, 
and the part
of the Lagrangian ${\bi A}\cdot\dot{\balpha}$ is replaced by ${\bi W}\cdot{\bomega}$, with
${\bi W}=\partial L/\partial{\bomega}$. In fact
\[
{\bi A}\cdot\dot{\balpha}\equiv\frac{\partial L}{\partial\dot\alpha_i}\dot\alpha_i=
\frac{\partial L}{\partial\omega_j}\frac{\partial\omega_j}{\partial\dot\alpha_i}\dot\alpha_i=\frac{\partial L}{\partial\omega_j}\omega_j\equiv
{\bi W}\cdot{\bomega}.
\]

The most general structure of the Lagrangian, in a relativistic and non-relativistic formalism, 
is thus
 \begin{equation}
L=T\dot{t}+{\bi R}\cdot \dot{\bi r}+{\bi V}\cdot\dot{\bi v}+{\bi W}\cdot{\bomega},
 \label{eq:gen2}
 \end{equation}
where $T=\partial L/\partial\dot{t}$,  ${\bi R}=\partial L/\partial\dot{\bi r}$,  ${\bi V}=\partial L/\partial\dot{\bi v}$, 
and ${\bi W}=\partial L/\partial{\bomega}$, as before, to be determined once the Lagrangian is given.
In general, since we postulate translation and rotation invariance, the Lagrangian will be independent
of $t$, ${\bi r}$ and $\balpha$.

Although we have not formulated yet
any particular Lagrangian it is possible to give some 
general features of the different systems in terms of the general
expression (\ref{eq:gen2}) and of these defined momenta. 
Noether's theorem, under the Galilei group transformations, 
gives rise to the ten following constants of the motion: \cite{Rivasl}

Under translations, the energy and linear momentum, respectively
 \begin{equation}
H=-T-{\bi v}\cdot\frac{d{\bi V}}{dt},\quad {\bi P}={\bi R}-\frac{d{\bi V}}{dt},
 \label{eq:HP}
 \end{equation}
under boosts the kinematical momentum
 \begin{equation}
{\bi K}=m{\bi r}-{\bi P}t-{\bi V},
 \label{eq:K}
 \end{equation}
and finally under rotations the angular momentum
 \begin{equation}
{\bi J}={\bi r}\times{\bi P}+{\bi v}\times{\bi V}+{\bi W}={\bi L}+{\bi S}.
 \label{eq:J}
 \end{equation}

The time derivative of the constant ${\bi K}$ leads to 
\[
{\bi P}=m{\bi v}-\frac{d{\bi V}}{dt},
\]
and compared with (\ref{eq:HP}) we get that ${\bi R}=m{\bi v}$, 
and the linear momentum of the system has not the direction of the velocity of the
point ${\bi r}$, provided the Lagrangian depends on the acceleration, and therefore the term ${\bi V}$
is nonvanishing. Point ${\bi r}$ does not represent the center of mass position of the particle. 
In fact if we define the position of a point ${\bi q}$ by
\[
{\bi q}={\bi r}-{\bi V}/m\equiv {\bi r}-{\bi k},
\]
then the kinematical momentum can be written as ${\bi K}=m{\bi q}-{\bi P}t$, and its time derivative
leads to ${\bi P}=md{\bi q}/dt$. Point ${\bi q}$, which is a different point than ${\bi r}$ whenever
${\bi V}$ is different from zero, represents the position of the centre of mass of the particle.
Vector ${\bi k}={\bi V}/m$, 
represents the relative position of point ${\bi r}$ with respect to the center of mass
${\bi q}$.

The angular momentum (\ref{eq:J}) contains two parts: One, ${\bi L}={\bi r}\times{\bi P}$, which 
is the orbital angular momentum of the system with respect to the origin, and another
${\bi S}={\bi v}\times{\bi V}+{\bi W}$, which is translation invariant and can be interpreted as the 
spin of the system. The orbital angular momentum is not a constant of the motion because 
${\bi P}$ is not pointing
along the velocity of point ${\bi r}$.
After taking the time derivative of the constant
angular momentum we get that the spin of a free particle satisfies the dynamical equation
 \begin{equation}
\frac{d{\bi S}}{dt}={\bi P}\times{\bi v},
 \label{eq:spindyn}
 \end{equation}
which is the same dynamical equation satisfied by the Dirac spin operator in the quantum formulation.
It is the classical spin observable equivalent to Dirac's spin operator.

We see that the structure of the spin is twofold. One ${\bi W}=\partial L/\partial\bomega$, 
coming from the dependence of the Lagrangian on the angular velocity, like in a rigid body, 
and we thus expect to be along the angular
velocity for a spherical symmetric object, and another ${\bi v}\times{\bi V}$, orthogonal to the velocity
of point ${\bi r}$, 
which can be written in terms of the relative position vector ${\bi k}$ as
\[
{\bi v}\times{\bi V}=-{\bi k}\times m{\bi v}.
\]
It looks like an (anti-)orbital part of the momentum $m{\bi v}$ located at ${\bi r}$ 
with respect to the center of mass.
The term anti-orbital comes because of the minus sign in front of it.

That point ${\bi r}$ represents the position of the charge can be seen when we consider
the interaction of the system with some external source. 
The most general interaction Lagrangian will be of the form
\[
L_I=-\phi\dot{t}+{\bi A}\cdot\dot{\bi r},
\]
{\it i.e.}, what is called the minimal coupling, independent of the variables 
$\dot{\bi v}$ and $\bomega$ because if the system is elementary
its spin cannot be modified by the interaction. The presence of terms in $\dot{\bi v}$ and $\bomega$
in the interaction Lagrangian,
will produce a modification of the functions ${\bi V}$ and ${\bi W}$ which define the spin. 
If we assume that the external potentials $\phi$
and ${\bi A}$ are only functions of $t$ and ${\bi r}$, 
then the external force is the Lorentz force where the fields are defined
at the position ${\bi r}$ and it is the velocity of point ${\bi r}$ which enters into the
magnetic force term.
 
\section{The Classical Relativistic Spinning Electron}

In the relativistic case we have three possible maximal, disjoint, 
homogeneous spaces of the Poincar\'e group,
spanned by the ten variables $(t,{\bi r},{\bi v},\balpha)$, 
provided $v$ is lesser, greater or equal to $c$. The very important
manifold for the description of the classical photon and electron is the one with $v=c$, because
it describes a system whose position vector ${\bi r}$, which is not the center of mass of the system,
it is moving at the speed of light. The importance of this manifold is suggested by Dirac's previous analysis
of the electron. In fact, it is the quantization of this system which leads to Dirac equation \cite{RivasQ}.

The Lagrangian has the same expansion in terms of the derivatives as in (\ref{eq:gen2}). 
For this system there is a small difference in the Noether constants of the motion,
with respect to the Galilei case. The expression for $H$, ${\bi P}$ and ${\bi J}$ is the same as 
in (\ref{eq:HP}) and (\ref{eq:J}). The ${\bi K}$ momentum is now given by
 \begin{equation}
{\bi K}=\frac{1}{c^2}H{\bi r}-{\bi P}t-\frac{1}{c^2}{\bi S}\times{\bi v},
 \label{eq:Krel}
 \end{equation}
with ${\bi S}={\bi v}\times{\bi V}+{\bi W}$, which satisfies 
the same dynamical equation as before (\ref{eq:spindyn}),
and is only a constant of the motion in the center of mass frame. 
The center of mass (or center of energy) of the electron is defined now as:
\[
{\bi q}={\bi r}-\frac{1}{H}{\bi S}\times{\bi v},
\]
because it leads to ${\bi P}=(H/c^2)d{\bi q}/dt$, by derivation of ${\bi K}$. 
Whenever the spin of the system is different from
zero the center of mass is a different point than ${\bi r}$, 
which, by the same arguments as before, 
represents the position of the charge. 

\begin{figure}
\centering
\includegraphics[width=.5\textwidth]{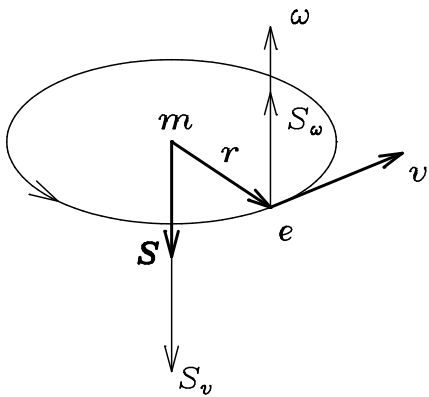}
\caption{\small{Evolution of point ${\bi r}$ in the center of mass frame. The body frame,
which rotates with angular velocity $\bomega$, is not depicted.}}
\label{fig1}
\end{figure}

For the center of mass observer ${\bi q}=0$, the spin is a constant of the motion, $H=mc^2$ and thus 
 \begin{equation}
{\bi r}=\frac{1}{mc^2}{\bi S}\times{\bi v},
 \label{eq:rS}
 \end{equation}
so that point ${\bi r}$ is moving in circles, 
at the speed of light, on a plane orthogonal to the constant vector ${\bi S}$. 
Classical mechanics does not restrict the value of the constant spin $S$ which 
can be any positive real number.
Its true value will be uniquely fixed after quantization.
The radius of this circle is $R=S/mc$ and the angular velocity of this internal 
motion or {\it zitterbewegung} is $\omega=mc^2/S$. By inspection of (\ref{eq:rS}) we see that the relative
orientation between the different magnitudes is the one depicted in fig.\ref{fig1}, in which we also
depict the two parts of the spin ${\bi S}_\omega={\bi W}$ along the angular velocity $\bomega$,
and the antiorbital ${\bi S}_v={\bi v}\times{\bi V}$, orthogonal to the zitterbewegung plane.
The antiparticle corresponds to the time reversed motion, or to consider that $H=-mc^2$.

The total spin ${\bi S}$ has the same direction as the antiorbital part so that the ${\bi S}_v$ part has to be larger
than the other ${\bi S}_\omega$. When quantizing the system the antiorbital part
only quantizes with integer values while the half integer comes from the quantization of the rotational part
${\bi S}_\omega$, in the opposite direction \cite{Rivasg}. 
This twofold structure of the spin leads to the classical concept 
of gyromagnetic ratio as will be shown below.

When we analyze the system in the center of mass frame and, therefore, 
three translational degrees of freedom are suppressed, the system reduces to a mechanical
system of only three degrees of freedom. These are the two coordinates $x$ and $y$ of the
point ${\bi r}$ on the zitterbewegung plane and the phase $\alpha$ of the rotation 
of the body frame which rotates with angular velocity $\bomega$. However this phase is the same as the phase 
of the orbital motion. Since the motion is at a constant velocity $c$, then the system
is reduced to a single degree of freedom, for instance, the $x$ coordinate. But as far as the $x$
coordinate is concerned its motion is a one-dimensional harmonic motion of frequency $\omega=mc^2/S$.
When we quantize this system, since it represents an elementary particle, it has no excited states
and therefore its allowed energy is just the ground state energy of the one-dimensional harmonic
oscillator $\hbar\omega/2=mc^2$ in this frame.
When compared with the classical obtained value of $\omega$, we get that after quantization the value of the 
classical spin parameter $S=\hbar/2$. This model of a classical spinning particle corresponds after quantization
to a fermion.

The classical expression that leads to Dirac equation when quantizing the system comes from (\ref{eq:Krel}).
If we take the time derivative of this constant of the motion and after that, the scalar product of the resulting
expression with ${\bi v}$ we get the classical equivalent of Dirac's Hamiltonian
\[
H={\bi P}\cdot{\bi v}+\frac{1}{c^2}{\bi S}\cdot\left(\frac{d{\bi v}}{dt}\times{\bi v}\right).
\]
This is a linear relationship between $H$ and ${\bi P}$, where the velocity ${\bi v}$ 
should be replaced by Dirac's velocity operator $c\balpha$
and the last term corresponds to $\beta mc^2$ in terms of Dirac's $\beta$ matrix \cite{RivasQ}.
In fact, the three vectors in the last term are orthogonal vectors. In the center of mass frame the absolute
value of the acceleration is $c^2/R$, so that taking into account the value of $R$ we get that this term
reduces to $\pm mc^2$, the positive value for the particle and the negative one for the antiparticle.

One can feel uncomfortable while talking about a massive particle whose position ${\bi r}$
is moving at the speed of light. We must remark that ${\bi r}$ does not represent the center of mass
or center of energy of the system. In our starting analysis, as the description of ${\bi r}$ 
as the origin of some inertial reference frame, we must add that we are not talking about reference frames moving at the speed
of light but rather about different inertial frames that we change from one to another at our will according 
to the evolution of the particle. 
The particular observers of our analysis could be, for instance, those observers which,
at every instant $\tau$ have the origin of its frame at the charge position, measure the velocity
of the charge along its $OX$ axis, the acceleration along the $OY$ axis, and thus the spin
of the electron along the $OZ$ axis.
All the kind of particles this formalism produces have a center of mass
at rest or moving with a velocity below $c$.

\section{Features of the Model}

If we analyze the classical model of the electron we see that the charge of the electron 
is at the point ${\bi r}$, 
but this point is moving at the speed of light in a motion known as the {\it zitterbewegung},
in a confined region of radius $\hbar/2mc$ around the center of mass, and oscillating with a 
frequency $2mc^2/h$. We see that the region of influence of this motion is Compton's wavelength. 
The charge is at a single point and this point like description 
is consistent with the latest LEP experiments
at CERN, which suggest that the charge has to be confined in a region below $10^{-19}$m, six orders
of magnitude smaller than Compton's wavelength.

Now in this kind of classical models, the electric charge is at a single point so that the problems 
related to a spatial charge distribution
are avoided. The charge is moving for every observer 
and therefore it creates a magnetic field. The electron is a point-like current, which is never at rest.
The charge motion produces
with respect to the center of mass a magnetic moment ${\bmu}$ 
in the direction orthogonal to the zitterbewegung plane
and which is related mechanically with the antiorbital part of the spin. 
But the spin has another part ${\bi S}_\omega$
related to the motion of the body frame in the opposite direction. 
When quantizing the system the total spin
is half the antiorbital part and this shows a plausible origin of the 
value $g=2$ for the gyromagnetic ratio  \cite{Rivasg}. 
But also, from the center of mass observer point of view,
there is an oscillating electric dipole moment ${\bi d}=e{\bi r}$.
We can say with Dirac that, in addition to the electric charge, 
{\it `the electron will therefore behave as 
though it has a magnetic moment $\bmu$ and an 
electric moment ${\bi d}$}.
The correspondence of this last observable with the quantum Dirac electric 
dipole moment is shown in \cite{Rivasl}. 
I think this is the electric dipole observable Dirac disliked.
It is oscillating at very high frequency and it basically plays no role in low energy 
electron interactions because its average value vanishes, 
but it is important in high energy physics or in very close electron-electron interactions.

For instance, if two electrons have their spins parallel, the corresponding electric dipoles
oscillate at the same rate and are contained in parallel planes. 
Therefore from the point of view of the two electrons
the electric dipoles always conserve the same relative orientation, and although
their average value is zero it is clear that the electric interaction between them
is not vanishing and in addition to the magnetic interaction this electric interaction
has to be taken into account. If the two particles are very close and the electric dipoles
have opposite orientation it is possible an attractive force.
It has been recently shown that this effect would produce,
from the classical viewpoint, the formation of metastable bound pairs of electrons
with parallel spins, when separated by a distance 
below Compton's wavelength and provided its relative velocity
is below some estimated value \cite{Rivaselec}.

\begin{figure}
\centering
\includegraphics[width=.5\textwidth]{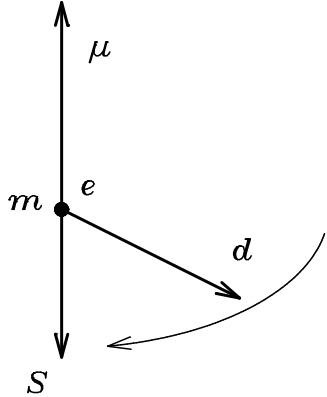}
\caption{\small{Approximate structure of the electron in the center of mass frame. It is a point of mass $m$
and negative charge $e$. 
It has a constant spin ${\bi S}$ and a constant magnetic moment $\bmu$ in the opposite direction,
and also an oscillating electric dipole moment ${\bi d}$, with frequency
$\omega=2mc^2/\hbar\simeq 10^{21}$ s$^{-1}$, in a plane orthogonal to ${\bi S}$.}}
\label{fig2}
\end{figure}

If we locate mathematically a positive and negative charge $\pm e$ in the center of mass we 
can have an approximate structure of the classical spinning electron as a point particle
of mass $m$ and charge $e$ in the center of mass of the system,
with the additional electromagnetic attributes $\bmu$ and ${\bi d}$ 
as the ones depicted in figure \ref{fig2}. 
The relative orientation 
between the spin and magnetic moment of the classical model 
depends on the sign of the charge of the object
considered as the particle while the antiparticle, which is described by the 
time reversed motion of the other and with opposite charge,
will have the same relative orientation between the spin and magnetic moment, because a current
in the reversed direction but of opposite charge will produce the same magnetic moment. 
It seems that the spin and magnetic moment
of the electron are opposite to each other, which corresponds to consider that the positive charged
object is the particle. 

\section{The Photon}

In the manifold spanned by the variables $(t,{\bi r},{\bi v},\balpha)$, 
with $v(\tau)=c$, we have two possibilities. Since the absolute value of the velocity is 
always constant,
then ${\bi v}\cdot\dot{\bi v}=0$. One possibility
is that $\dot{\bi v}=0$, the particle moves in straight lines at the speed of light. 
We have in this case the description of a photon.
The other is that $\dot{\bi v}\neq0$ but always orthogonal to the velocity. This possibility
leads to the electron description, as has been shown above.

For the photon, since $\dot{\bi v}=0$, there will be no $\dot{\bi v}$ term in the 
expansion of the Lagrangian. It is a system of six degrees of freedom. Three represent the position
of a point and the other three its orientation in space and the 
spin will be related to the change of orientation
but not to the dependence of the Lagrangian on the acceleration. 
The particle moves and rotates, but the Lagrangian is not
a function of the acceleration, which vanishes.
The homogeneity condition implies that the Lagrangian 
for a photon is
\[
L=\epsilon\frac{S}{c}\frac{\dot{\bi r}\cdot\bomega}{\dot{t}},
\]
where $\epsilon=\pm1$ will be interpreted as the helicity and $S$ the absolute value of the spin, which
from the classical point of view is unrestricted.
The spin is now
\[
{\bi S}=\frac{\partial L}{\partial\bomega}=\epsilon S\frac{{\bi v}}{c},
\]
which is a vector parallel ($\epsilon=+1$) or antiparallel  ($\epsilon=-1$) to the velocity
of propagation, and it is never transversal to the motion. Its value is independent of the observer,
and thus a nonvanishing intrinsic property.

The linear momentum is
\[
{\bi P}=\frac{\partial L}{\partial\dot{\bi r}}=\epsilon\frac{S}{c}\,\bomega,
\]
in a time evolution description ($\dot{t}=1$). Since the spin is constant 
and $d{\bi S}/dt={\bi P}\times{\bi v}=0$, this means that ${\bi P}$,
and thus $\bomega$, lie along ${\bi v}$.

The energy of the photon is
\[
H=-\frac{\partial L}{\partial\dot{t}}={\bi S}\cdot\bomega, 
\]
and if it is definite positive both ${\bi S}$ 
and $\bomega$ have the same direction. 

Because the spin structure is related to the rotation variables it can quantize 
with all integer and half integer values \cite{Rivasl}. If, 
when quantized we take $S=\hbar$ and thus 
$H=\hbar\omega=h\nu$. The frequency of the photon
is the frequency of its rotation around the direction of motion, leftwards or rightwards
according to the spin orientation.

Because $H^2-{\bi P}^2c^2=0$, the photon is a massless system.

If, when quantized, we take for $S=\hbar/2$, we are describing both left and right handed 
neutrinos, but this classical formalism does not discriminate the left handed ones.

\section{Conclusions}

We have seen that by describing the evolution of the consecutive updated inertial observers,
which measure the elementary particle in the same kinematical state, we can describe the states and
dynamics of an elementary particle. Therefore we use the variables which characterize every inertial
observer as the classical variables which define the kinematical state of the particle. We have found 
a clear interpretation between the motion of the charge at ${\bi r}$ at the speed of light and
the quantum mechanical Dirac analysis of the electron. Both, the magnetic and electric dipole moments,
have had a clear classical explanation. 

Nevertheless, the classical model has several classical inconsistencies. The stationary motion
of the charge is accelerated and therefore the system has to loose energy by radiation. As an alternative, 
we see that although the charge is accelerated, the center of mass of the free electron is not and 
thus there is no variation of the kinetic energy of the free particle. 
It is suggesting that, from the classical
viewpoint, we must revisit the classical theory of 
the radiation of spinning objects and associate
the energy radiated with the acceleration of the center of mass. Now the model
is more complex because we have to describe the evolution of two points ${\bi r}$ and ${\bi q}$. Alternatively,
we can only describe the evolution of the center of charge ${\bi r}$, 
but it satisfies then a fourth order differential equation, which is more difficult to analyze.

The electric field produced by this model of spinning electron it is neither static nor Coulomb like. 
Nevertheless its time average value during a 
complete turn of the charge produces an electrostatic field which is Coulomb like from a distance of
three Compton's wavelength from the origin to infinity and which does not diverge at ${\bi r}=0$, 
but it vanishes there.
The time average of the magnetic field produces a static 
magnetic field which corresponds to the magnetic field
produced by a magnetic moment $\bmu$ at the origin. 
This static magnetic field does not diverge at the origin
\cite{Rivasl}. In this time average there is no electric dipole and this justifies the usual picture
of the electron as a charged object with a magnetic moment and no electric dipole. All these fields
have some infinities which go like $1/r$, and not like $1/r^2$, 
in some points of the zitterbewegung plane.

This picture predicts that the spin and magnetic moment of the particle and antiparticle
have the same relative orientation. 
Nevertheless it is argued that for the electron they are antiparallel
while they are parallel for the positron. To our knowledge, 
no clear experimental evidence of this relative orientation
can be found in the experimental literature. Most of the very 
accurate measurements of the electron magnetic moment
and of the $g-2$ anomaly are done by analyzing the precession 
frequency of the spin in external magnetic fields. But this precession 
frequency is independent of whether spin and magnetic moment 
are parallel or antiparallel vectors. 
Two plausible experiments for this relative measurement
have been recently proposed \cite{Rivasmu}.

This formalism is still at a seminal level, but some of the features it is able to describe
are very promising. The usual spinless physics is contained within it. But the interest is that we
can handle classical dynamical equations for spinning systems. With the use of these dynamical 
equations we have been able to show that there exists a nonvanishing classical probability
of tunneling when the electron is properly polarized \cite{Rivastunnel}, and also that polarized
electrons can form bound states of spin 1, and thus a gas of bosons, provided they are very close
and with no very high relative velocity \cite{Rivaselec}. 
Spin 1 electron pairs seem to be the most probably state of condensed electrons
in ferromagnetic superconductors.

\section*{Acknowlegdments}

This work has been supported by the Universidad del 
Pa\'{\i}s Vasco/Euskal Herriko Unibertsitatea research grant 9/UPV00172.310-14456/2002.

\end{document}